\documentclass[oldversion]{aa}
\usepackage{txfonts}
\usepackage{subfigure}
\usepackage{graphicx}
\def\lax    {\ifmmode{_<\atop^{\sim}}\else{${_<\atop^{\sim}}$}\fi}
\def\gax    {\ifmmode{_>\atop^{\sim}}\else{${_>\atop^{\sim}}$}\fi}
\def\kms    {\ifmmode{{\rm ~km\,s}^{-1}}\else{~km\,s$^{-1}$}\fi}

\def\approx   {$\sim$}

\def\arcmper  {\ifmmode \rlap.{' }\else $\rlap{.}' $\fi}

\def\arcsper  {\ifmmode \rlap.{'' }\else $\rlap{.}'' $\fi}
\def\arcsgper  {\ifmmode \rlap.^{s }\else $\rlap{.}^s $\fi}

\def\deg      {\ifmmode^\circ\else$^\circ$\fi}     

\def\hper     {\ifmmode \rlap.^{h}\else $\rlap{.}^h$\fi}

\def\m1       {$^{-1}$}

\def\mper     {\ifmmode \buildrel m\over . \else $\buildrel m\over .$\fi}

\def\to       {\rightarrow}

\def\>           {$>$}
\def\<           {$<$}

\def\simlt       {\lower.5ex\hbox{$\; \buildrel < \over \sim \;$}}
\def\simgt       {\lower.5ex\hbox{$\; \buildrel > \over \sim \;$}}

\begin{document}
\title{The AMIGA sample of isolated galaxies} 
\subtitle{II. Morphological refinement
\thanks{This research has made use of the  LEDA 
(http://leda.univ-lyon1.fr) database and the NASA/IPAC 
Extragalactic Database (NED) which is 
operated by the Jet Propulsion Laboratory, California Institute of Technology, 
under contract with the National Aeronautics and Space Administration. 
This work is partially based on observations made with the 1.5 m telescope 
of the Observatorio de Sierra Nevada, Granada, Spain, which is operated 
by the IAA (CSIC). }\fnmsep\thanks{Full tables~\ref{OSN}, \ref{rev-morfo}, 
\ref{morfo-bib} and \ref{morfo-osn} are available in electronic form
at the CDS via anonymous ftp to {\tt cdsarc.u-strasbg.fr} ({\tt 130.79.128.5}) 
or via {\tt http://cdsweb.u-strasbg.fr/cgi-bin/qcat?J/A+A/}}}
\author{J. W. Sulentic\inst{1}
\and
L. Verdes-Montenegro\inst{2}
\and
G. Bergond\inst{2,3}
\and
U. Lisenfeld\inst{4}
\and
A. Durbala\inst{1}
\and
D.~Espada\inst{2}
\and
E.~Garcia\inst{2}
\and
S.~Leon\inst{2}
\and
J.~Sabater\inst{2}
\and
S. Verley\inst{2,5}
\and
V. Casanova\inst{2}
\and
A. Sota\inst{2}}
\offprints{L.\,Verdes-Montenegro, email:\,{\tt lourdes@iaa.es}}
\institute{Department of Astronomy, University of Alabama, Tuscaloosa, USA
\and Instituto de Astrof\'{\i}sica de Andaluc\'{\i}a, CSIC,
Apdo. 3004, 18080 Granada, Spain
\and GEPI, Observatoire de Paris--Meudon, 61 Avenue de l'Observatoire,
75014 Paris, France
\and Departamento de F\'\i sica Te\'orica y del Cosmos, 
Facultad de Ciencias, Universidad de Granada, Spain
\and LERMA, Observatoire de Paris, 61 Avenue de l'Observatoire,
75014 Paris, France}
\date{Received / accepted }

\abstract{We present a complete refinement of the optical morphologies for 
galaxies in the Catalog of Isolated Galaxies (Karachentseva 
\cite{karachentseva73}) that forms the basis of the AMIGA (Analysis of the
interstellar Medium of Isolated GAlaxies) project. 
Uniform reclassification using the digitized  POSS\,II benefited from the high 
resolution and dynamic range of  that sky survey. 
Comparison with independent classifications made for an SDSS 
overlap sample of more than 200 galaxies confirms the reliability of the 
early vs. late-type discrimination and the accuracy of spiral subtypes 
within $\Delta T = $ 1--2. CCD images
taken at the Observatorio de Sierra Nevada were also used to solve 
ambiguities in early versus late-type classifications.
A considerable number of galaxies in the catalog 
($n =$ 193) are flagged for the presence of  nearby companions or signs of 
distortion likely due to interaction. 
This most isolated sample of galaxies in the local  Universe is dominated by 
two populations: 1) 82\% spirals (Sa--Sd) with the bulk being luminous 
systems with small bulges (63\% between types Sb--Sc) and 2) a significant 
population of early-type E--S0 galaxies (14\%). Most of the types later than 
Sd are low luminosity galaxies concentrated in the local supercluster where 
isolation is difficult to evaluate. The late-type spiral majority of the 
sample spans a luminosity range $M_{B{\rm -corr}} = -$18 to $-$22 mag. 
Few of the 
E/S0 population are more luminous than $-$21.0 marking an absence of, an often 
sought, super $L^*$ merger (e.g. fossil elliptical) population. The rarity of 
high luminosity  systems results in a fainter derived $M^*$ for this 
population compared to the spiral optical luminosity function (OLF). 
The E--S0 population is from 0.2 to 0.6 mag
fainter depending how the sample is defined.  This marks the AMIGA sample as 
almost unique among samples that compare  early and late-type OLFs separately. 
In other samples, which always involve galaxies in  higher density 
environments, $M^*_{\rm E/S0}$ is  almost always 0.3--0.5 mag brighter than 
$M^*_{\rm S}$, presumably reflecting  a stronger correlation between $M^*$ and 
environmental density for early-type galaxies.
 }
\maketitle

\section{Introduction\label{sect1}}

The AMIGA project (Analysis of the interstellar Medium of Isolated GAlaxies, 
see {\tt http://www.iaa.es/AMIGA.html})
involves identification and parameterization of a statistically significant
sample of the most isolated galaxies in the local Universe. Our goal is to
quantify the properties of different phases of the interstellar media of
the galaxies least likely affected by their external environment. In an 
earlier paper (Verdes-Montenegro et al. \cite{verdes05}; hereafter Paper I) we 
summarized the optical properties of the Catalog of Isolated Galaxies
(CIG) and presented an improved 
OLF. That work showed that CIG is a 
reasonably complete sample ($\sim$80\%) down to m$_{B{\rm -corr}} \sim 15.0$
and within $\sim$100~Mpc. Analysis of the redshift and magnitude distributions 
suggests that CIG  ($n =$ 1050) can be interpreted in five parts:
\begin{enumerate}
\item  A local supercluster population ($n \sim 150$) rich in
dwarf galaxies  (within $V_{\rm R} \sim 1500$ km\,s$^{-1}$)
and largely unsampled in the rest of the CIG which involves
galaxies with radial velocities
$V_{\rm R} =$ 1500--15\,000 km\,s$^{-1}$. While many
are regarded as members of groups within the local supercluster,
some have been noted for their isolation (e.g. CIG 45, Makarova
\& Karachentsev \cite{makarova98}; CIG 121, Karachentsev et
al. \cite{karachentsev96}; CIG 524, Uson \& Matthews \cite{uson03};
CIG 624, Drozdovsky \& Karachentsev \cite{drozdovsky00}).

\item  A local supercluster population within  $V_{\rm R} \sim 3000$ 
km\,s$^{-1}$ that contributes a few more luminous ($M_{B{\rm -corr}} < -19$) 
and possibly isolated galaxies to the CIG ($n \sim 50$).

\item   A contribution from the lowest surface density parts of the
Pisces-Perseus supercluster in the range $V_{\rm R} =$ 4000--6000 
km\,s$^{-1}$ ($n \sim 100$).

\item   A quasi-homogeneous population of isolated galaxies that
account for about 50\% of the total sample within $V_{\rm R} =$ 10\,000 
km\,s$^{-1}$. This  contribution is as close to a ``field population'' 
as exists in the local Universe. Early claims for such a component in 
a largely independent sample (Turner \& Gott \cite{turner75}) were 
later challenged (Huchra \& Thuan \cite{huchra77} (14 CIG in their 
sample of 39); Haynes \& Giovanelli \cite{haynes83}) ($n \sim 500$).

\item  The  remaining 250 CIG galaxies lie mostly between $V_{\rm R} =$ 
10\,000--15\,000 km\,s$^{-1}$ forming a high redshift tail to 
quasi-homogeneous component 4) and involving some of the most luminous objects 
in the sample. Inclusion/exclusion from an OLF calculation will only affect 
the bright end.
\end{enumerate}

The main goal of this paper is to present a revision of optical morphologies 
for the CIG  based upon the POSS\,II images. All of the above components are
included in the revision in order to facilitate creation of well-defined
subsamples later on. An ancillary goal involves identification of certain 
and suspected  examples of CIG galaxies involved in one-on-one interaction. 
A comparison is made between previous classifications as well as recent 
results for restricted samples based on our own CCD data as well as  
Sloan Digital Sky Survey (SDSS) images. Finally 
we  present  type-specific 
OLFs and compare them with other, mostly recent, OLF derivations.

\section{Past work on CIG morphologies\label{sect2}}

The first attempts at morphological revision of the CIG began in the
years immediately after its publication (Karachentsev \&
Karachentseva \cite{karachentsev75}; Arakelyan \cite{arakelyan84}).
They were hampered by the lack of any significant number
of images superior to those of POSS\,I. The POSS\,I-based early and late-type
populations for the CIG are 168 (E/S0) and 883 (S/I) respectively
(Sulentic \cite{sulentic89}). Despite their low resolution, classifications
from POSS\,I based on the Kodak 103a emulsions were at least uniform.
A radio continuum survey of the CIG (Adams et al. \cite{adams80})
also provided an upgrade of POSS\,I classifications using the glass plates
rather than the photographic print version. Image-tube data for 64 
likely or possible early-type galaxies were also provided. They assigned
E/S0 and spiral  classifications to 120 and 440 galaxies, respectively, in
the approximately  half of the CIG that they examined. An \ion{H}{i} survey 
of a bright subsample of the CIG showed evidence that at least some of the 
early-type
galaxies in the CIG were misclassified (Haynes \& Giovanelli \cite{haynes84}).
Attempts to isolate the early-type fraction in CIG have continued to this
day (Aars et al. \cite{aars01}; Saucedo-Morales \& Beiging \cite{saucedo01};
Stocke et al. \cite{stocke04}).

A few detailed studies of CIG galaxies, recognized as very isolated,
also exist (CIG 947, Verdes-Montenegro et al. \cite{verdes95}; CIG 121,
Karachentsev et  al. \cite{karachentsev96}; CIG 710, Verdes-Montenegro
et al. \cite{verdes97}; CIG 164, 412, 425, 557, 684, 792, 824, 870, 877,
Marcum et al. \cite{marcum04}; CIG 96, Espada et  al. \cite{espada05}).
CIG have also been included in many detailed studies of smaller samples 
of isolated galaxies (number of CIG galaxies follows each reference): 
Xanthopoulos \& de Robertis \cite{xanthopoulos91} (1);
Marquez \& Moles \cite{marquez96}, \cite{marquez99} (4);
Morgan et al. \cite{morgan98} (3); Aguerri \cite{aguerri99} (6);
Colbert et al. \cite{colbert01} (1); Kornreich et al. \cite{kornreich01} (1);
Pisano et al. \cite{pisano02} (4); Madore et al. \cite{madore04} (1);
Reda et al. \cite{reda04} (2).

The goal of the AMIGA project is to extract a
significant subsample of the {\em most} isolated galaxies from the CIG which
should be the most isolated galaxies in the local Universe. 
The need for a large and uniformly
selected sample of isolated galaxies is obvious especially if one wants
to evaluate morphology as a function of isolation. The size of the CIG
sample allows one to refine and yet retain a sample large enough to distinguish
degrees of isolation and morphology statistical. Our morphology refinement
complements the upcoming refinement (Verley et al. 2005) of  probabilistic
isolation by identifying close pairs and peculiar galaxies that might
remain undetected in that more automated study.

\section{The data\label{sect3}}

The only available classifications for a majority of AMIGA/CIG 
galaxies come from POSS\,I and this motivated our uniform survey with 
POSS\,II. POSS\,I was based on the 103a--O and E emulsions providing 
broad-band blue and red images for all CIG galaxies. POSS\,II is based
on the  IIIa--J and --F emulsions which provide higher contrast and 
resolution  ($\sim$100 lines/mm vs. $\sim$\,60 lines/mm for 103a emulsions). 
The higher contrast (dynamic range) is especially important for 
recognizing spiral galaxies with a high surface brightness bulge embedded 
in a lower surface brightness disk. Overall the higher resolution and 
contrast POSS\,II improved discrimination between E/S0 and spiral subtypes
as well as detection of close companions. The SDSS provided CCD images 
for 215 CIG galaxies through  the 3rd Data Release (DR3). SDSS $r$ band 
images (scale 0\farcs396/pixel) were extracted  from the SDSS archive 
for all of the CIG overlap sample. This comparison sample offers
an excellent test of the reliability of POSS\,II results. 

We have obtained CCD images for more than 120 CIG galaxies with 
the 1.5 m Sierra Nevada Observatory (OSN) telescope near 
Granada (Spain). The observations involve galaxies for which
POSS\,II classifications are regarded as uncertain. We used
a 2k$\times$2k EEV CCD camera giving a $\sim$\,8$'\,\times8'$ 
field with 0\farcs23 pixels. Most images were obtained in 
the $V$ and $R$ or $V$ and $I$ bands, and reduced using standard IRAF
tasks\footnote{IRAF is distributed by
the National Optical Astronomy Observatory, which is operated by AURA, Inc.,
under cooperative agreement with the National Science Foundation.}.
Table~\ref{OSN} lists all CIGs observed at OSN with the explicited filters 
and exposure times, as well as the seeing. 

\begin{table}
\caption{CIGs observed with the OSN 1.5-m telescope CCD camera.
Filters, exposure times and seeing are indicated$^{1}$.}
\label{OSN}
\centering
\begin{tabular}{l l l l }
\hline
\hline
CIG & Filters & Exposures (s)& Seeing \\ 
\hline
3&$VRI$&1800/1800/1800&1\farcs8/1\farcs8/1\farcs8\\
8&$VI$&900/900&2\farcs0/2\farcs0\\
14&$VR$&1800/1800&1\farcs6/1\farcs4\\
21&$VI$&900/900&1\farcs7/1\farcs7\\
23&$VI$&900/900&2\farcs3/1\farcs8\\
\ldots & \ldots& \ldots & \ldots\\
\hline
\end{tabular}\\
$^1$ The full table is available in electronic
form  from  CDS or at {\tt http://www.iaa.csic.es/AMIGA.html}.
\end{table}

\section{POSS\,II morphologies for the CIG\label{sect4}}
\subsection{Classification considerations}

\begin{figure*} 
\resizebox{17.9cm}{!}{\rotatebox{0}{\includegraphics{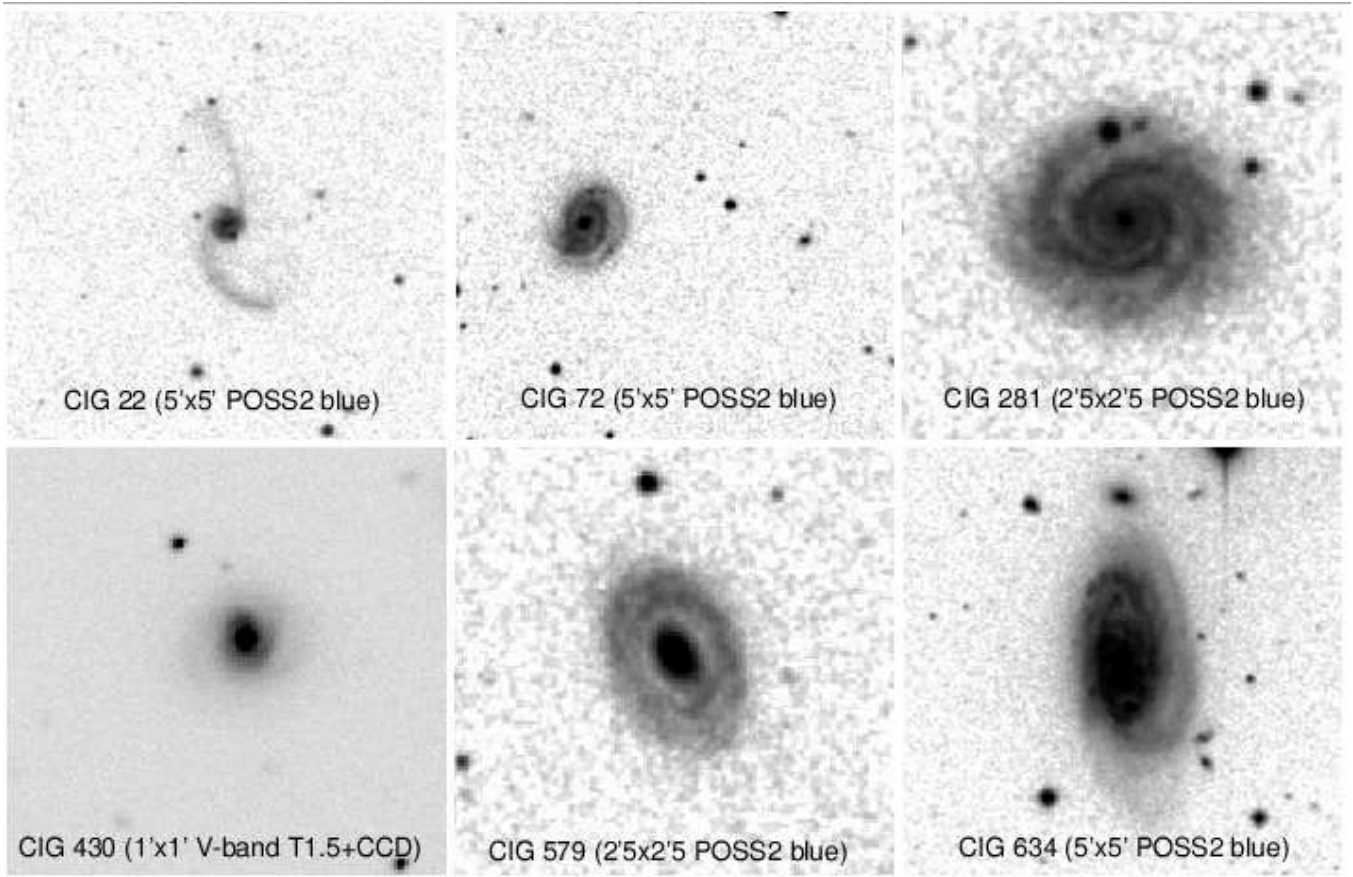}}}
\caption{Six examples of CIG galaxies from POSS\,II images unless otherwise  
noted. {\it Upper left}: CIG 22, classified I/A=y, shows ``integral sign'' 
structure almost certainly due to interaction although a companion cannot
yet be identified.  {\it Upper centre}: CIG 72, an Sc spiral showing both disk
distortion and an active nucleus with a likely dwarf companion. {\it Upper
right}: CIG 281, a prototype isolated Sc spiral. {\it Lower left}: for CIG 
430, the OSN CCD images (stack of 3$\times$900 s exposures in $V$ shown 
here, seeing 1\farcs4) reveal faint  
spiral arms in a CIG often previously classified as early-type. {\it Lower
centre}: CIG 579, a rare prototype isolated galaxy with a large bulge, 
classified as Sab. 
{\it Lower right}: CIG 634, classified I/A=y showing morphological 
distortion and a LINER nucleus; dwarf companion visible on the northern edge.}
\label{imag}
\end{figure*}

POSS\,II images were evaluated (see Sect.~\ref{sect3}) using {\tt ds9} which
enabled us to control zoom and  scaling functions while deriving morphological
types. Figure~\ref{imag} shows six examples of CIG galaxies that illustrate 
examples of 
specific types or  problems. Each image is labeled with a CIG designation as 
well as information about the origin of the image. The individual images will 
be discussed in the text where appropriate. 
All types were  derived by JWS with AD providing independent estimates
for the $\sim$\,200 galaxies previously classified as early-type.
Classifications for the bulk of the sample follow the basic Hubble sequence
with spiral sub-types estimated from the observed bulge to disk ratio
(e.g. $B/D\sim 0.5$ = Sb). In the majority of CIG spirals this ratio 
is reasonably unambiguous, however for some spirals the presence 
of an inner ring can confuse the
classification. A small nuclear bulge (e.g. $B/D \sim 0.25$) indicates an
Sc type but a small bulge embedded in an inner ring can cause one to assign
an earlier type. In some cases a small nuclear bulge can be resolved within
the ring and other times not. These galaxies would be classified Sc and
Sab--Sb, respectively. Only detailed surface photometric
studies can  resolve this kind of ambiguity. The images for CIG 281 and 579 
show, respectively, examples of common small (Sc) and rare large (Sab) bulge 
spirals. Openness of spiral arms is not taken as a
type indicator but rather as an indication of tidal perturbation. We argue,
for example, that it is meaningless to assign a standard Hubble type of Sc
or Sb to a galaxy like CIG 22 (Fig.~\ref{imag}).  While the presence/absence 
of a bar was noted in unambiguous cases the results are unlikely to carry much 
statistical weight given the plate-to-plate variations at POSS\,II resolution.
Another difficulty  involves distinguishing between E, S0 and Sa types. 
We found the IIIa images to be surprisingly effective for detecting a 
disk component (via an inflection
in $I(r)$) in early types. This means that differentiating between E and S0 was
effective with discrimination between S0 and Sa the larger challenge especially
beyond $V_{\rm R} \sim 10\,000$ km\,s$^{-1}$.

\subsection{Distorted morphologies/minor interactions}

We removed $n = 33$ objects from the sample that unambiguously violate
the goal of the CIG catalog. Thirty-two involve interacting  systems
(CIG 6, 22, 31, 62, 63, 76, 80, 85, 126, 146, 247, 293, 349, 439, 468, 634,
687, 701, 761, 773, 787, 796, 809, 819, 853, 921, 940, 946, 977, 1027 and 1038)
 with the other identified as a local globular cluster (CIG 781). 
We also  noted $n =$ 161 entries where 
interaction is suspected based upon evidence for
asymmetries/distortions that might be of tidal origin, as e.g. CIG 72.
CIG 72 and 634 (with companions) are illustrated in Fig.~\ref{imag}. 
In the latter 
case the companion redshift is consistent with physical association while in 
the former the companion has not been identified. Some
of these objects were noted in  previous CIG analysis but we have chosen to 
start from scratch in order to proceed in an uniform way.
  
Statistics in Table~\ref{numbers} are based upon the 
1018 CIG sources that remain when the clearly interacting galaxies are removed.
This tabulation does not exhaust the number of CIG with faint ``companions''
because some evidence of perturbation of the CIG primary was required to
warrant designation as suspected interaction. Surprisingly, many of the 
latter cases 
show no evidence for a companion brighter than $-$17 mag. A fundamental
question raised by this result involves how much detectable 
kinematic/morphological
perturbation can be caused by a dwarf companion (see Espada et
al. \cite{espada05} for the beginning of detailed follow-up on some of these
cases). This statement assumes of course that the features were not produced
by an accretion event which, in any case, is more difficult to prove.
All of the suspected interacting  systems were assigned a Hubble  type and 
retained in the statistics. Many/most additional, especially late-type 
spirals, are accompanied by small low surface brightness objects that in many 
cases could be dwarf companions in the $-$15 absolute magnitude range. Their 
POSS\,II detectability depends on the quality of a particular plate. 
Perhaps, in this sense, no luminous spiral can be called ``isolated''.

\begin{table}
\caption{Results of the morphological
reevaluation of the CIG sample (for $V_{\rm R}$ $>$ 1000 km\,s$^{-1}$).}
\label{numbers}
\begin{center}
\begin{tabular}{lrclrrc}
\hline\hline
Type&$T$&$n$& $n/1018$&I/A=?&$n_{\rm SDSS}$& $n_{\rm SDSS}$/215\\
\hline
I/A    &    &  32 &  ---    &      0  &        &          \\
E    & $-$5 & 58 &  0.057  &      1  &    7   &   0.032 \\
E/S0 & $-$3 & 14 &  0.014  &      0  &    4   &   0.019 \\
S0   & $-$2 & 67 &  0.066  &      3  &    17  &   0.079 \\
S0/a &  0 & 19 &  0.019  &      2  &    7   &   0.033\\
Sa   &  1 & 13 &  0.013  &      2  &    3   &   0.014\\
Sab  &  2 & 52 &  0.051  &      8  &    11  &   0.051\\
Sb   &  3 & 159&  0.156  &      20 &    25  &   0.116\\
Sbc  &  4 & 200&  0.196  &      40 &    33  &   0.153 \\
Sc   &  5 & 278&  0.273  &      68 &    69  &   0.321\\
Scd  &  6 & 61 &  0.060  &      7  &    15  &   0.070\\
Sd   &  7 & 41 &  0.040  &      7  &    13  &   0.060\\
Sdm  &  8 & 15 &  0.015  &      0  &    7   &   0.033\\
Sm   &  9 & 15 &  0.015  &      1  &    3   &   0.014\\
Im   &  10& 26 &  0.026  &      2  &    7   &   0.033\\
\hline
E--S0  &  &139&  0.137   &       4   &   28   &  0.130\\
Sa--Sd &  &804&  0.790   &     152   &   169  &  0.786 \\
Sb--Sc  &  &637&  0.626   &     128   &   127  &  0.591 \\
 \hline
\end{tabular}
\end{center}
\end{table}

\subsection{Results of the reclassification}

Table~\ref{rev-morfo} presents results of the POSS\,II based morphological
reevaluation for $V_{\rm R}\,>$\,1000 km\,s$^{-1}$ and is formatted as 
follows: 1) CIG number,
2) estimated Hubble type (a ``:'' indicates uncertain type and need for
better imaging data), 3) a ``y'' indicates secure presence 
and a ``n'' clear absence of a bar, 
4) a ``y'' indicates a morphologically
distorted system and/or almost  certain interacting system
while "?" indicates evidence for interaction/asymmetry with/without certain
detection of a companion. Table~\ref{morfo-bib} presents a tabulation of 
literature classifications
for CIG galaxies within  $V_{\rm R} = 1000$ km\,s$^{-1}$. 
This table is a companion to Table~2 in Paper I that summarized 
redshift independent distance determinations for these nearby galaxies.
Galaxies within $V_{\rm R} <$ 1000 km\,s$^{-1}$ are tabulated separately
because standard Hubble morphologies are not very useful for these local 
galaxies. Table~\ref{numbers} summarizes the breakdown of Hubble subtypes
in terms of the number and sample fraction. The numerical scale is taken from
RC3 and the correspondences are given in the table.

\begin{table}
\caption{New morphologies for the $V_{\rm R} > 1000$ km\,s$^{-1}$ CIG 
sample$^{1}$}
\label{rev-morfo}
\begin{center}
\begin{tabular}{llcc}
\hline
\hline
CIG & $T$(RC3) & Bar&  Interacting \\
\hline
      1&          5  &   n   &      ?     \\ 
                    2&          5  &   y    &            \\ 
                    3&          3: &           &            \\     
                    4&          5  &          &            \\ 
                    5&          4  &           &            \\ 
                   \ldots&          \ldots&     \ldots&   \ldots\\
\hline
\end{tabular}
\end{center}
$^1$ The full table is available in electronic
form  from  CDS or at {\tt http://www.iaa.csic.es/AMIGA.html}.
\end{table}

\begin{table}
\caption{Compiled morphologies for the $V_{\rm R}\!<\!1000$ km\,s$^{-1}$ CIG 
sample$^1$.}
\label{morfo-bib}
\begin{center}
\begin{tabular}{llc}
\hline\hline
CIG & Morphology & Reference$^2$\\
\hline
  45&   Im/BCD   &  1\\
  ''  & SAm      &  2\\
 105&   SAB(s)d  &  3\\
  ''&   SBc(s)   &  4\\
  ''&   SBc      &  5\\
 \ldots&   \ldots      &\ldots\\
\hline
\end{tabular}
\end{center}
$^1$ The full table is available in electronic
form  from  CDS or at {\tt http://www.iaa.csic.es/AMIGA.html}.\newline
$^2$ (1) NED, (2) van Zee (\cite{vanzee01}), (3) Baggett et al. 
(\cite{baggett98}), (4) CAG (Sandage \& Bedke \cite{sandage94}),
(5) Burda \& Feitzinger (\cite{burda92}), \ldots
\end{table}
 
Table~\ref{numbers} reveals that the CIG is dominated by two types of 
galaxies: 1) late-type galaxies with 82\% of the CIG in the range Sa--Sm 
($T = $ 1--9) and 2) early-type E--S0 galaxies comprising about 14\% of the 
sample. Early-type spirals are quite rare with Sa--Sab representing only 6\% 
of the sample while Sb--Sc are the prototype CIG population compromising 63\%. 
The early-type spiral fraction may be even smaller than the numbers suggest 
as some of these have very uncertain classifications. Distinction between types
Sa/Sab/Sb is more  ambiguous than for Sb/Sbc/Sc 
and a large part of this ambiguity involves the more frequent
presence of inner rings in the former range. Given the uncertainties 
about degree of isolation  for types later than $T = $ 6--7 that are: 
a) largely within the local supercluster and b) undetectable beyond a few 
1000 km\,s$^{-1}$, we are unable to characterize any very isolated low 
luminosity population.

\begin{figure}
\resizebox{8.5cm}{!}{\rotatebox{0}{\includegraphics{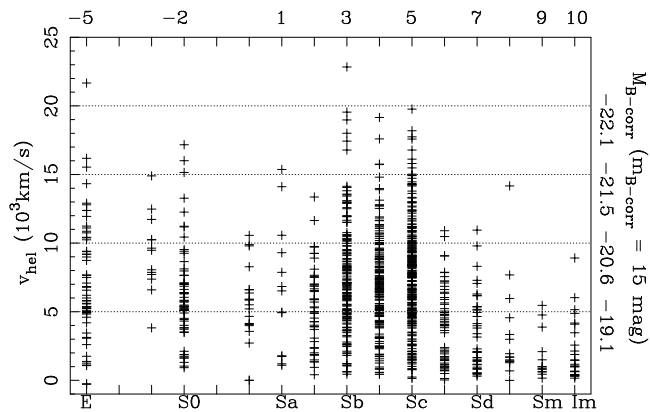}}}
\caption{Distributions of new morphologies as a function of 
recession velocity ($V_{\rm R}$). The right ordinate indicates the 
absolute magnitude for a $m_{B{\rm -corr}}$ = 15.0 galaxy
using H$_0$ = 75 km\,s$^{-1}$\,Mpc$^{-1}$.}
\label{fig-t-v}
\end{figure}

\begin{figure}
\resizebox{8.5cm}{!}{\rotatebox{0}{\includegraphics{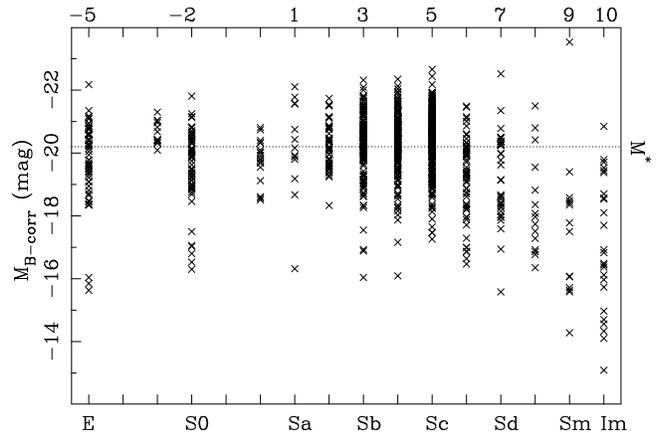}}}
\caption{Distributions of new morphologies as a function of absolute 
magnitude $M_{B{\rm -corr}}$.
The abscissa indicates basic Hubble subtypes ({\it bottom})
 and corresponding RC3 numeral types ({\it top}). The  
dotted line shows the total sample $M^*$ derived in Paper I.}
\label{fig-morfo-Mabs}
\end{figure}

Figure~\ref{fig-t-v} shows the distributions of new morphologies as a function 
of recession velocity ($V_{\rm R}$) while Fig.~\ref{fig-morfo-Mabs} shows 
the distributions as a  function of absolute magnitude ($M_{B{\rm -corr}}$). 
The latter are derived from  $m_{B{\rm -corr}}$
(Paper I) assuming H$_0 = 75$ km\,s$^{-1}$\,Mpc$^{-1}$. The horizontal
dotted line indicates the sample $M^*$ derived in Paper I for the most
complete part  of the CIG between  m$_{B{\rm -corr}}$ = 11 and 15.0 mag.

\begin{figure*} 
\resizebox{16cm}{!}{\rotatebox{0}{\includegraphics{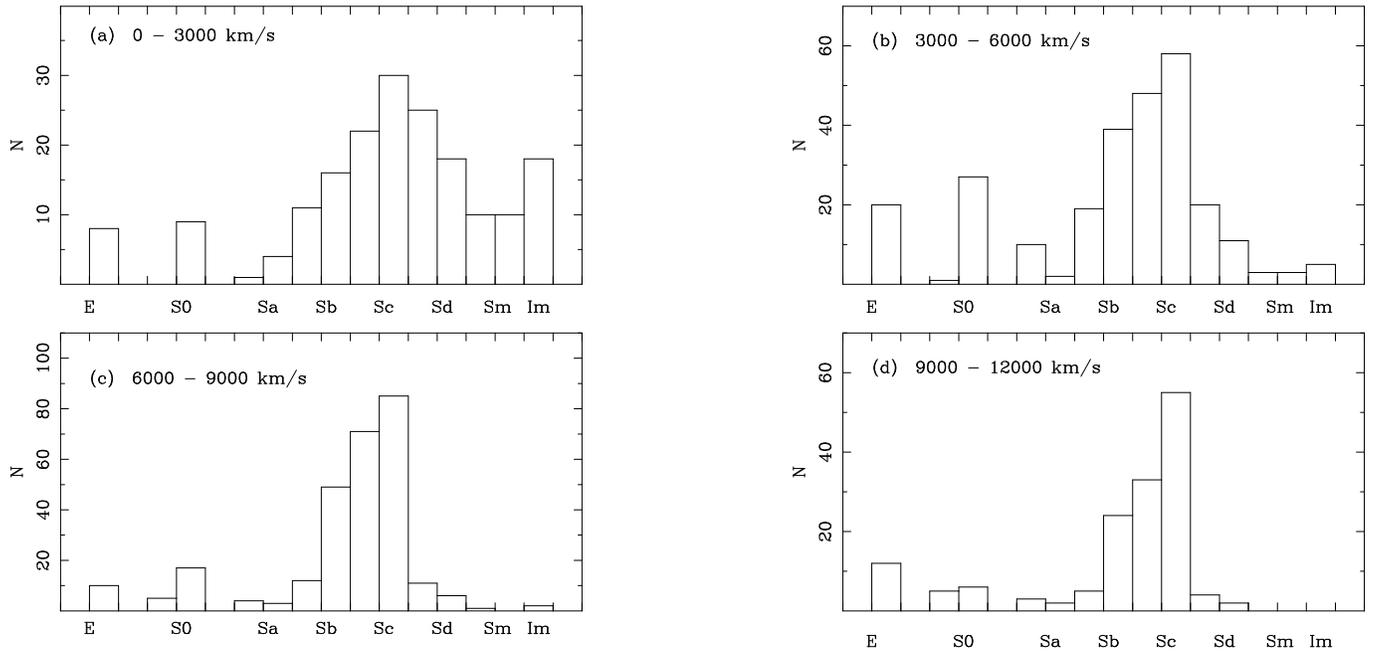}}}
\caption{Type distribution in four recession
velocity bins following the velocity breakdown  used in Fig.~2 of Paper I.}
\label{fig-t-fdv}
\end{figure*}

\subsection{POSS\,II versus old and new classifications}   
\begin{figure*} 
\resizebox{16.5cm}{!}{\rotatebox{0}{\includegraphics{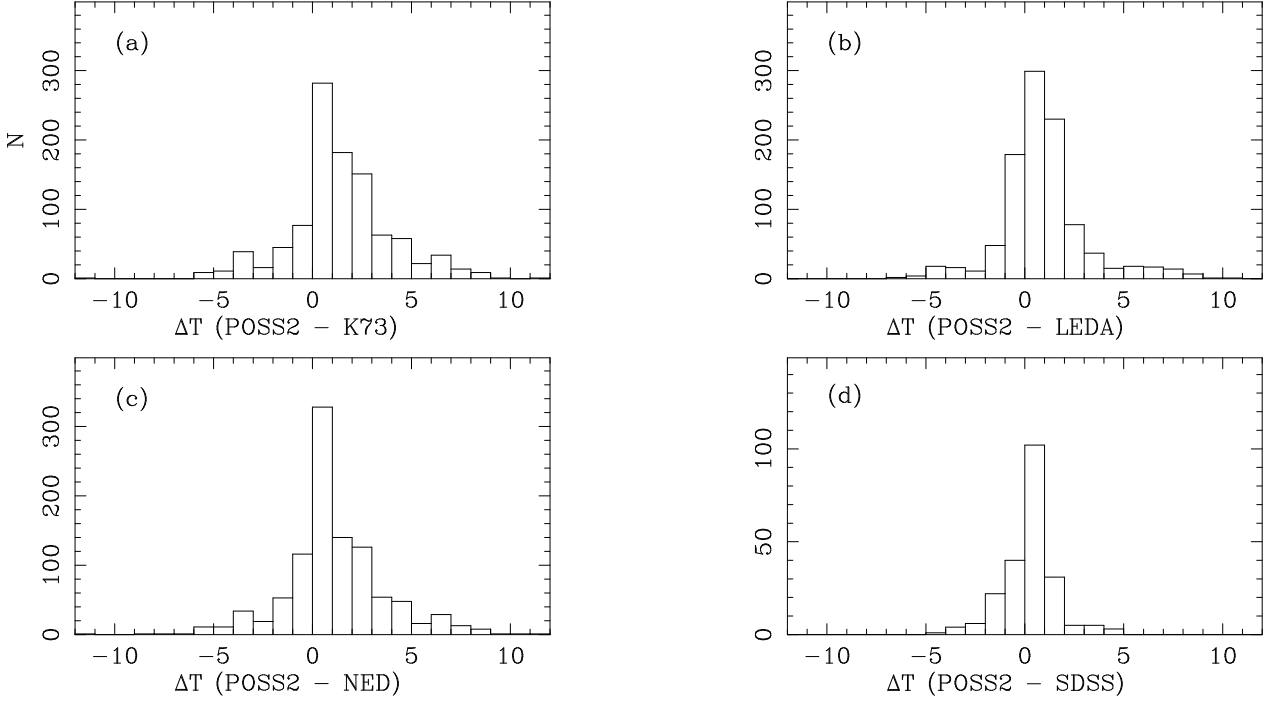}}}
\caption{Comparison of our new classifications with those from: (a) the 
original CIG compilation (Karachentseva \cite{karachentseva73}, K73) based 
upon POSS\,I, (b) the LEDA (c) the NED and (d) the CCD images of the SDSS.}
\label{fig-dt}
\end{figure*}

Comparison of CIG types from the literature for individual
CIG galaxies sometimes spans the  entire Hubble sequence. 
POSS\,II classifications represent an order of magnitude improvement
in reliability especially for: 1) discrimination between early (E--S0) and
late-types (Sa--Im)  as well as 2) determination of spiral subtypes
to within $\Delta T = 1$--2. Figure~\ref{fig-dt} compares our new 
classifications with those from: a) the original CIG compilation (Karachentseva
\cite{karachentseva73}, K73) based upon POSS\,I, b) the Lyon-Meudon 
Extragalatic Database (LEDA), c) the NASA/IPAC Extragalactic Database (NED) 
and d) the CCD images of the SDSS. LEDA and NED types represent a partial 
improvement over POSS\,I because they include literature types
based on higher quality photographic or electronic images. At least half 
of the  sample  types remain POSS\,I after the upgrades provided by 
these somewhat redundant  samples. Figure~\ref{fig-dt} presents histograms 
showing the distribution of differences in type assignment in the 
sense POSS\,II $-$ ``other'', 
where ``other'' can be  K73, LEDA, NED or SDSS. The RC3 system contains no 
major type designation for $-4$ or $-1$ so for the purpose of Fig.~\ref{fig-dt}
the early types were moved to E = $-$3, E/S0 = $-$2 and S0 = $-$1, 
in order to maintain a 
constant interval between all types in the comparison histograms.
Two trends are reflected in the K73, LEDA and NED comparisons: 1) an
asymmetry  favoring small positive differences, and 2) a very broad base 
with values from $-6$ to $+9$. The former shows the tendency for spiral types 
to become later because bulges are better defined with POSS\,II. The latter 
reflects larger changes from early- to late-type or vice versa.
The LEDA database appears to give the most reliable measures prior to POSS\,II.

A good test of the robustness of POSS\,II classification comes from an overlap
sample of $n =$ 215 CIG galaxies in SDSS. This CCD-based survey provides
the best available seeing limited images. We reevaluated Hubble types
using the SDSS images without reference to the POSS\,II
classifications. Figure~\ref{fig-dt}d shows the results
of a comparison in the sense POSS\,II $-$ SDSS. Since the same observer made
both sets of classifications this will be a test of consistency/robustness
rather than absolute accuracy of type assignments.
The results of this comparison are encouraging in the sense 
that there is close agreement between POSS\,II and SDSS derived types.
We find exact agreement ($\Delta T = 0$) for about one half of the
overlap sample probably reflecting the ease of recognition of the
majority Sb--Sc population on both datasets. The bulk of additional objects 
($n =$ 65) lie within $\Delta T =\pm1$. There is a slight asymmetry
toward negative values, which reflects the ability of SDSS to resolve
the bulge component in spirals more easily than POSS\,II, resulting in a
shift toward later type.

SDSS confirms the core populations identified with POSS\,II. 58\% of the 
galaxies in the SDSS subsample are concentrated in the range Sb--Sc. 
The mostly local late-type (Sd--Im) contributes about 14\% (same as POSS\,II)
while Sa--Sab galaxies contribute about 6\%. 
The E--S0 fraction in SDSS is similar to POSS\,II at 
14\% with S0 apparently twice as numerous as ellipticals, representing 
a decrease in the number of E and an increase in the S0 population.  
12 of 39 objects in the SDSS subsample previously assigned 
 I/A=? (see Table~\ref{numbers})
are rejected. Either signs of distortion were not confirmed on the SDSS images 
or available SDSS spectra revealed that suspected companions showed a 
different redshift from the CIG galaxy.  There are thirty  CIG galaxies 
whose type was changed from E--S0 to spiral in our POSS\,II  reevaluation. 
Most of these  objects did not show obvious spiral structure on POSS\,II. 
However in 
our judgment they showed colors (blue) and structure (flatter than $R^{1/4}$ 
law luminosity distribution or evidence for high spatial frequency structure) 
more similar to distant spirals than to E/S0 systems.  All six galaxies from 
this population with SDSS data confirm our spiral classifications. SDSS 
confirms the utility of POSS\,II for galaxy classification in the local 
Universe.

Table~\ref{morfo-osn} lists morphologies
derived from our new observations from the OSN, as well as types
from POSS\,II to facilitate comparison, together with the interaction status.
An additional 69 galaxies observed did not warrant a type change either
because of confirmation or because the new CCD data was obtained in
seeing conditions which did not improve upon the POSS\,II discrimination.
Most changes were small except for a population of early-type crossovers.
The new CCD data suggest a change from late to early type for 13 galaxies
(3 to E, 1 to E/S0 and 9 to S0) and a change from early to late-type for 5
galaxies. Not surprisingly most of the crossovers involve the faintest
galaxies in CIG ($n =$ 10 with $m_{B{\rm -corr}}$ $>$ 15.0). 
The remaining brighter galaxy
changes ($n =$ 8) that could  affect the OLF involve 4 early-type losses and
4 gains that effectively cancel out any possible change. These are the
most difficult objects to classify in our sample.

\begin{table}
\caption{Revised CCD morphologies from the OSN CCD database$^{1,2}$.}
\label{morfo-osn}
\begin{center}
\begin{tabular}{lrcrc}
\hline\hline
CIG&  \multicolumn{2}{c}{OSN}  & \multicolumn{2}{c}{POSS\,II}   \\  
 & $T$ & I/A &$T$ & I/A \\
\hline
21    &     &  ?       &  & \\
57    & $-$2  &          & 3  & \\
70    & $-$2  &          & 10  & \\
74    &  5  &    ?     & 4 & \\        
87    &     &          &   & ? \\
\ldots   & \ldots  &  \ldots    & \ldots  & \ldots\\
\hline
\end{tabular}
\end{center}
$1$ In 69 other cases no change was warranted
due to confirmation or images that did not improve upon POSS\,II.\newline
$^2$ The full table is available in electronic form 
from CDS or at {\tt http://www.iaa.csic.es/AMIGA.html}.
\end{table}

\begin{figure}
\resizebox{8.5cm}{!}{\rotatebox{0}{\includegraphics{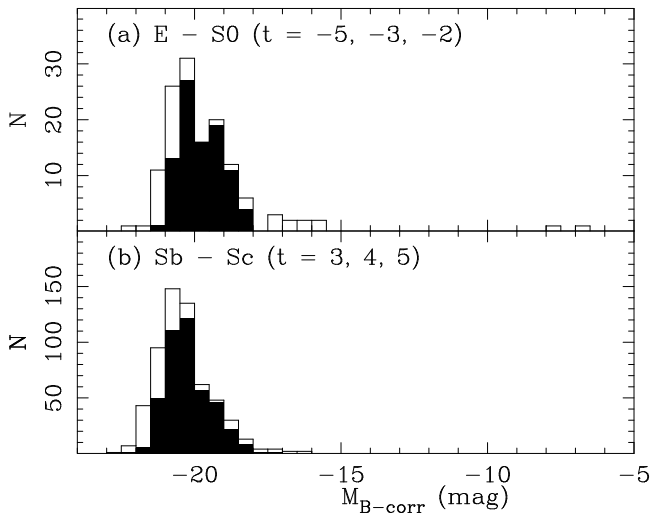}}}
\caption{Luminosity distribution for the indicated 
populations both for the complete sample and (solid) for galaxies in the range 
$V_{\rm R} =$ 2000--10\,000  km\,s$^{-1}$.}
\label{Mabs-t}
\end{figure}

\section{The OLF of isolated galaxies as a 
function of morphology\label{sect5}}

\subsection{Results for the CIG sample}

\begin{figure*}
\resizebox{17.cm}{!}{\rotatebox{0}{\includegraphics{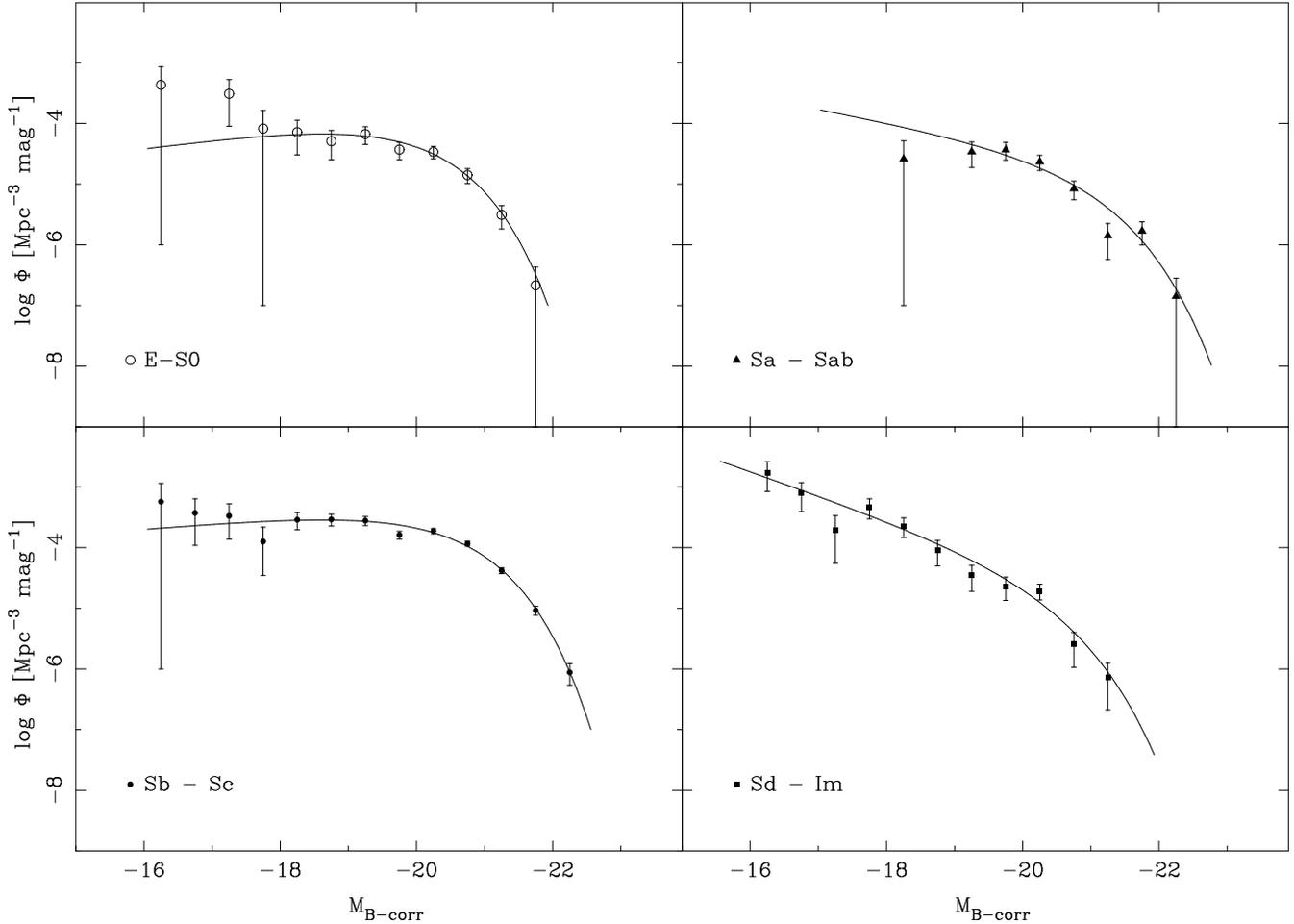}}}
\caption{OLF for the different morphological types present in the CIG sample
together with the corresponding  Schechter fit shown as a solid line.}
\label{fig-olf-morfo}
\end{figure*}

\begin{figure}
\resizebox{8.5cm}{!}{\rotatebox{0}{\includegraphics{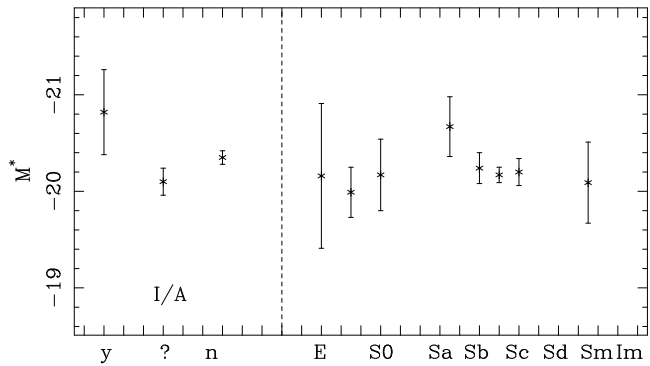}}}
\caption{The Schechter function $M^*$ parameter for the CIG 
sample as a function
of the morphological type.}
\label{m-type-cig}
\end{figure}

\begin{figure}
\resizebox{8.5cm}{!}{\rotatebox{0}{\includegraphics{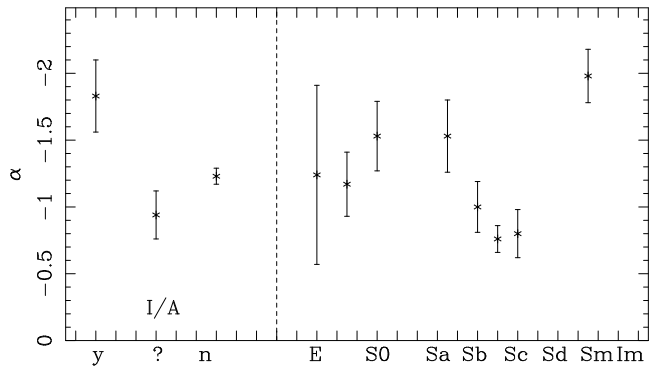}}}
\caption{The same as in Fig.~\ref{m-type-cig} for the $\alpha$ parameter.}
\label{alpha-type-cig}
\end{figure}

Figure~\ref{fig-olf-morfo}  shows OLFs for four morphological 
bins: E--S0, Sa--Sab, Sb--Sc, and Scd--Im. The first and
third bins (E--S0 and  Sb--Sc) are most important from the 
point of view of AMIGA. Table~\ref{olf}  gives best fit Schechter
function parmeters for the above bins as well as some extra binnings 
to facilitate comparison with other samples. The $M^*$ and $\alpha$ 
parameters for the morphological bins are plotted in Figs.~\ref{m-type-cig}
and \ref{alpha-type-cig}. The Sb--Sc population shows a best 
fit  $M^*$ that is 0.2--0.6 brighter than for the E--S0 population, 
confirming the scarcity of luminous early-types seen in Fig.~\ref{Mabs-t}. 
The E--S0 sample is rather small ($n = 139$ for the full sample and $n = 71$ 
restricting to galaxies in the range $11 \le m_{B{\rm -corr}} \le 15$). 
The result  is that the Schechter fit $M^*$ parameters  change with small 
sample modifications. In contrast the Sb--Sc OLF is much more robust with a 
sample of $n = 470$ galaxies in the  $11 \le m_{B{\rm -corr}} \le 15$ 
range. If we restrict the E--S0 derivation to  bins containing $n>4$  or 
more galaxies we obtain $M^* = -19.5$, yielding a difference of 0.6 between 
the E--S0 and Sb--Sc populations, rather than 0.2 mag obtained from 
Table~\ref{olf}. No matter how we cut our sample, 
the early-types are fainter than the Sb--Sc sample. In order to better 
assess the  significance of the difference between these samples we performed 
 a nonparametric characterization. Table~\ref{nonpar} 
shows the results of a comparison of the means and 25th, 50th and 75th 
for the  E--S0 and Sb--Sc populations, ranging from the full CIG subsamples to 
those corresponding to the most complete range 
(between $11\le m_{B{\rm -corr}}\le15$).
All of them confirm the fact that our E--S0 population is underluminous 
compared to our Sb--Sc population. 

The Sa--Sab population shows the brightest $M^*$ value for a Hubble type bin 
but with  correspondingly large uncertainty reflecting the small sample size. 
The $\alpha$ parameter  here and elsewhere must 
be considered of limited value in the absence of galaxies fainter than $-18$ 
to $-19$ in the bulk of our sample. The steep $\alpha$  parameter for the 
Scd--Im bin reflects the strong contribution of low luminosity dwarfs in the 
local 
part of the sample. The high value for $M^*$ for these late-types reflects 
the rigidity of the Schechter function which, in the presence of such a 
strong dwarf contribution yields an  artificially high $M^*$. This is 
amplified by bright Scd--Sd spirals (see Fig.~\ref{fig-morfo-Mabs}) included 
in that bin. The I/A=y subsample 
shows the brightest value of $M^*$ possibly reflecting two effects:
1) in some cases magnitude estimates may represent the combined light and 
2) optical luminosities in paired galaxies are enhanced by a factor of 
about 2.0 (Xu \& Sulentic \cite{xu91}). We see here evidence for the nurture 
signature that we are trying to avoid.  We note that there is no detectable
difference in the OLFs for galaxies with the I/A=? designation and the 
non-interacting ones.

\begin{table}
\caption{OLF for the CIG sample.}
\label{olf}
\begin{tabular}{lccccr}
\hline\hline
Types &  $\Phi$ (Mpc$^{-3}$\,mag$^{-1}$) & $\alpha$& $M^*$& N\\
\hline
E   &                        3.2($\pm$ 3.4)  $\times$ 10$^{-5}$& $-$1.24  $\pm$ 0.67& $-$20.16 $\pm$ 0.75 & 27\\
S0 &   4.0($\pm$ 2.6)  $\times$ 10$^{-5}$& $-$1.53  $\pm$ 0.26& $-$20.17 $\pm$ 0.37 & 36 \\
Sb &                    1.2($\pm$ 0.3) $\times$ 10$^{-4}$& $-$1.00  $\pm$ 0.19& $-$20.24 $\pm$ 0.16& 115 \\
Sbc     &              1.5($\pm$ 0.3) $\times$ 10$^{-4}$& $-$0.91  $\pm$ 0.17& $-$20.30 $\pm$ 0.14& 159 \\
Sc  &                 2.1($\pm$ 0.4) $\times$ 10$^{-4}$& $-$0.80  $\pm$ 0.18& $-$20.20 $\pm$ 0.14 & 196 \\
Sd  &                0.9($\pm$ 0.7) $\times$ 10$^{-4}$& $-$0.51  $\pm$ 0.39& $-$19.64 $\pm$ 0.46& 7\\

E--S0&                1.0($\pm$ 0.4)  $\times$ 10$^{-4}$& $-$1.17  $\pm$ 0.24& $-$19.99 $\pm$ 0.26 & 71 \\
Sa--Sab &                    3.0($\pm$ 1.6) $\times$ 10$^{-5}$& $-$1.53  $\pm$ 0.27& $-$20.67 $\pm$ 0.31 & 51 \\
Sb--Sc &                    5.1($\pm$ 0.5) $\times$ 10$^{-4}$& $-$0.76  $\pm$ 0.10& $-$20.17 $\pm$ 0.08 & 470\\
Sd--Im &                    4.5($\pm$ 0.4) $\times$ 10$^{-4}$& $-$1.98  $\pm$ 0.20& $-$20.09 $\pm$ 0.42 & 59\\
I/A=y  &            1.1($\pm$ 0.9) $\times$ 10$^{-5}$& $-$1.83  $\pm$ 0.27& $-$20.82 $\pm$ 0.44 & 24\\
I/A=? &                    1.6($\pm$ 0.3) $\times$ 10$^{-5}$& $-$0.94  $\pm$ 0.18& $-$20.10 $\pm$ 0.14& 132\\
I/A=n &                    6.1($\pm$ 0.7) $\times$ 10$^{-4}$& $-$1.23  $\pm$ 0.06& $-$20.35 $\pm$ 0.07 & 713\\
\hline
\end{tabular}
\end{table}
\begin{table*}
\caption{Non parametric statistics of $M^*$ for CIG subsamples}
\label{nonpar}
\begin{tabular}{lccccc}
\hline\hline
Sample &  75th & 50th&25th& Mean \\
\hline
E--S0: All                                     &  $-$20.593&   $-$20.080 &  $-$19.200 &$-$19.599 $\pm$ 0.167\\
E--S0: $V_{\rm R} > 1500$ km\,s$^{-1}$                    &  $-$20.625&   $-$20.140 &  $-$19.355 &$-$19.949 $\pm$ 0.086\\
E--S0: $V_{\rm R} > 1500$ km\,s$^{-1}$, $11\le m_{B{\rm -corr}}\le 15$ &  $-$20.622&   $-$20.230 &  $-$19.504 &$-$20.044 $\pm$ 0.106\\
Sb--Sc: All                                    &  $-$20.999&   $-$20.495 &  $-$19.925 &$-$20.355  $\pm$ 0.039\\
Sb--Sc: $V_{\rm R} > 1500$ km\,s$^{-1}$               &  $-$21.020&   $-$20.526 &  $-$19.984 &$-$20.418  $\pm$ 0.037\\
Sb--Sc: $V_{\rm R} > 1500$ km\,s$^{-1}$, $11\le m_{B{\rm -corr}}\le 15$&$-$21.072&   $-$20.590 &  $-$20.065 &$-$20.480  $\pm$ 0.041\\
\hline
\end{tabular}
\end{table*}

\subsection{Comparison of morphology dependent CIG OLF with other samples}

In this section we  compare our type-specific OLFs with results from other 
samples involving a range of environments.  There are two reasons for 
such a comparison: 1) OLFs as a function of type and environment have 
recently become available for several large galaxy surveys and 2) we want 
to see if our conclusions about the morphologies of galaxies in the
lowest density environments are consistent with these new survey results.
We concentrate on the shape of the OLF rather than the space density ($\Phi$) 
which is difficult to compare and of less interest in the AMIGA context. 
Derived Schechter fit parameters for all relevant subsamples are  detailed 
in Table~\ref{olf-env}. 
All published values  of $M^*$ have been  reduced to 
H$_0$ = 75 km\,s$^{-1}$\,Mpc$^{-1}$ and transformed to $m_{B{\rm -corr}}$ 
using the relations given in Paper I. This forms the principal basis for 
comparison because $\alpha$ is sensitive to the faint end 
which AMIGA cannot effectively sample beyond the local supercluster.  
\begin{table*}
\caption{OLF as a function of morphology and environment.}
\begin{tabular}{llrcc}
\hline\hline
Morphology&Sample &  $\Phi$ (Mpc$^{-3}$\,mag$^{-1}$) & $\alpha$&  M$_{B{\rm -corr}}$ \\
\hline
Early types& 2dFGRS - void1&   0.28($\pm$ 0.10)  $\times$ 10$^{-3}$   & $-$0.15  $\pm$ 0.53& $-$19.30 $\pm$ 0.33\\
Early types& 2dFGRS - void2&   0.68($\pm$ 0.17)  $\times$ 10$^{-3}$   & $-$0.43  $\pm$ 0.24& $-$19.84 $\pm$ 0.14\\
Early types& 2dFGRS - mean&   1.73($\pm$ 0.19)  $\times$ 10$^{-3}$   & $-$0.39  $\pm$ 0.11& $-$20.06 $\pm$ 0.08\\
Early types& 2dFGRS - cluster&  15.6($\pm$ 7.7)  $\times$ 10$^{-3}$   & $-$1.12  $\pm$ 0.14& $-$20.81 $\pm$ 0.18\\
Late  types& 2dFGRS - void1&   1.02($\pm$ 0.55)  $\times$ 10$^{-3}$   & $-$1.14  $\pm$ 0.24& $-$19.46 $\pm$ 0.19\\
Late  types& 2dFGRS - mean&   3.37($\pm$ 0.61)  $\times$ 10$^{-3}$   & $-$1.00  $\pm$ 0.07& $-$19.92 $\pm$ 0.07\\
Late  types& 2dFGRS - cluster&  22.7($\pm$ 12.2)  $\times$ 10$^{-3}$   & $-$1.09  $\pm$ 0.20& $-$20.02 $\pm$ 0.18\\
\hline
E--S0  & SSRS2 &  1.9($\pm$ 0.8)  $\times$ 10$^{-3}$  & $-$1.00  $\pm$ 0.09& $-$20.27 $\pm$ 0.10  \\
Sa--Sd  & SSRS2 &  3.4($\pm$ 1.4)  $\times$ 10$^{-3}$&  $-$1.11  $\pm$ 0.07& $-$20.33 $\pm$ 0.08\\
Irr--Pec &SSRS2 & 0.2($\pm$ 0.08)  $\times$ 10$^{-3}$ & $-$1.81  $\pm$ 0.24& $-$20.68 $\pm$ 0.50\\
\hline
E& NOG     & 0.46($\pm$ 0.12)  $\times$ 10$^{-3}$ & $-$0.47 $\pm$ 0.22 & $-$20.61 $\pm$ 0.26 \\
S0& NOG    & 0.81($\pm$ 0.20)  $\times$ 10$^{-3}$ & $-$1.17 $\pm$ 0.20 & $-$20.30 $\pm$ 0.26 \\
Sa--Sb& NOG      & 2.20($\pm$ 0.46)  $\times$ 10$^{-3}$ & $-$0.62 $\pm$ 0.11 & $-$20.37 $\pm$ 0.12 \\
Sc--Sd& NOG     & 3.12 ($\pm$ 0.59) $\times$ 10$^{-3}$  & $-$0.84 $\pm$ 0.10 & $-$20.25 $\pm$ 0.11 \\
Sm---Im& NOG     & 0.07($\pm$ 0.07)  $\times$ 10$^{-3}$ & $-$2.41 $\pm$ 0.28 & $-$20.97 $\pm$ 0.72 \\
E--S0& NOG     &  1.03($\pm$ 0.29)  $\times$ 10$^{-3}$   & $-$0.97  $\pm$ 0.14& $-$20.55 $\pm$ 0.18\\
Sa--Im & NOG   & 4.58($\pm$  0.73)  $\times$ 10$^{-3}$ & $-$1.10  $\pm$ 0.07& $-$20.49 $\pm$ 0.09\\
Field& NOG     &   & $-$1.10  $\pm$ 0.06& $-$20.53 $\pm$ 0.08\\
Groups& NOG  &   & $-$1.19  $\pm$ 0.10& $-$20.45 $\pm$ 0.12\\
\hline
& &   & & $M^*_r$ \\
\hline
All & SDSS - void&   0.08($\pm$ 0.04)  $\times$ 10$^{-2}$   & $-$1.18  $\pm$ 0.13& $-$20.36 $\pm$ 0.11\\
All&  SDSS - wall&   0.60($\pm$ 0.03)  $\times$ 10$^{-2}$   & $-$1.19  $\pm$ 0.07& $-$21.24 $\pm$ 0.08\\
\hline
\end{tabular}
\label{olf-env}
\end{table*}
Comparison samples include:
\begin{itemize}
\item  2dFGRS samples the redshift range $0.05 < z < 0.13$ down to 
$b_{\rm J} \sim 19.45$ and includes $n =$ 81387 galaxies (Croton et al. 
\cite{croton05}). It is divided between early- and late-types  based on 
spectral characteristics of the galaxies. They distinguish between ``void'', 
``mean'' and ``cluster'' environments in seven subsamples based on  the 
density contrast in spheres of radius $R = 8$ Mpc, with the  most extreme void 
subsample similar in size to the CIG. The CIG contains few galaxies in
recognized voids however  it is not clear that 2dFGRS makes any distinction 
between a void galaxy and a single galaxy that is very isolated. 
\item  The Second Southern Sky Redshift Survey (SSRS2, Marzke et al. 
\cite{marzke98}) 
samples a volume similar to CIG ($z<0.05$) down to a similar magnitude 
limit $m_{\rm SSRS2} = 15.5$, and contains $n =$ 5404 galaxies.  Morphological 
classifications come from several sources, ranging from detailed to rough 
designations. Three broad morphological classes are defined: E/S0, spiral 
and irregular/peculiar, without environmental distinction. 
\item The Nearby Optical Galaxy (NOG) sample (Marinoni et al.
\cite{marinoni99})
involves $n =$ 6392 galaxies within  $V_{\rm R}$ = 5500 km\,s$^{-1}$ and 
brighter than $B = 14.0$, therefore corresponding to the inner part of CIG. 
They distinguish subsamples according to various group properties  
(Garcia \cite{garcia93}) for a total of 4025 galaxies.  Any galaxy not 
included in one of the group categories is considered ``field''. The 
morphologies were compiled by Garcia et al. (\cite{garciaetal93}) from RC3. 
\item A sample of 1000 ``void'' galaxies extracted from the early data 
release (sample 10 in Blanton et al. \cite{blanton03} involves 155126 
galaxies) of SDSS (Rojas et al. \cite{rojas04}, Hoyle et al. \cite{hoyle05}).  
This volume limited sample extends out to $z = 0.089$. 
The void sample is compared to a ``wall'' sample of 
12732 galaxies drawn from the same database. The void sample spans a galaxy
luminosity range that is similar to CIG.
\item A sample of 102 E and S0 galaxies from the CIG studied by Stocke et 
al. (\cite{stocke04}) after morphological revision based on new images for 
80 and 86\% of the E and S0 galaxies, respectively. The remaining galaxies  
in the sample were given POSS\,I classifications. 
\end{itemize}

Croton et al. (\cite{croton05}) find the 2dFGRS void population to be composed 
of primarily late-type galaxies with early-types dominating the cluster 
population. Early-types  are also seen in the void sample which is consistent 
with our result where 14\%  of AMIGA galaxies show E--S0 morphologies. The 
2dFGRS early-type OLF shows an $M^*$ that decreases systematically (by 1.5 
magnitudes) from cluster to void environments. Comparison of Tables~\ref{olf} 
and \ref{olf-env} shows that our early-type $M^*$ value is close to their 
``void2'' population. If we remove from our OLF derivation the sample bins 
with $n<4$  galaxies as described above our results are closest to the 
2dFGRS ``void1'' result (within $\Delta M^* \sim 0.2$) more consistent with 
AMIGA representing, as we argue, the most extreme local isolated galaxy 
sample. The dependence of $M^*$ for early-types on the  local environment is 
also found in the  analysis of an SDSS sample  (Hogg et al. \cite{hogg03}),
where ``red'' galaxies are found to be sensitive to environmental overdensity.
 
The situation for late-types is less clear. In the SDSS sample ``blue''
galaxies are relatively insensitive to the environment, while late-type 
(blue) galaxies  in the 2dFGRS show little change in the OLF across all  
density environments except for  a luminosity decrease in the void populations.
  However our $M^*$ value for late-types is from 0.6--1.1 magnitudes 
brighter than the 2dFGRS ``void1'' value. One possible explanation for this 
difference is that 2dFGRS includes a large population of low-medium luminosity 
isolated spirals that  we do not sample. If so it is well disguised (at least 
brighter than  $M_{B{\rm -corr}}$  $\sim -17$) because their late-type 
$\alpha$ parameter values are only slightly steeper than ours.
We note that their late-type void $M^*$ values are the lowest of any we 
consider in Table~\ref{olf-env} including an SDSS void estimate discussed 
later. Hence we interpret the better agreement of CIG with the 2dFGRS 
early-type OLF to indicate that the disagreement in the late-type 
population involves an underestimate of $M^*$ for the 2dFGRS. 
 Our late-type $M^*$ is consistently brighter than our early-type value 
and this is confirmed by more robust tests of the difference in mean 
luminosity between our early and late-type samples. Overluminous spirals 
are common in our sample while overluminous early-types are rare. 
If a bias was operating in the CIG selection process then one would expect 
it to favor the overluminous early-types at the expenses of the less 
luminous galaxies.

SSRS2 finds no significant difference between the OLFs for early- and 
late-type (Sa--Sd) galaxies. 
Our late-type $M^*$ value agrees closely with  theirs while our early-type 
sample is fainter than their  corresponding value. This is most likely due 
to the lack of environmental discrimination in SSRS2 which permits 
inclusion of overluminous E--S0 that  prefer denser environments. 
It is difficult to compare $\alpha$ because SSRS2 goes deeper than 
AMIGA and should more effectively 
sample the  dwarf galaxy population. Our Sd--Im bin shows a faint end $\alpha$ 
parameter similar to the Irr/Pec bin of SSRS2 again perhaps reflecting the  
strong  contribution from local dwarfs that is present in our sample. 
The bright $M^*$ must be attributed to the ``peculiar'' galaxy part of 
that population. 

The NOG sample (Marinoni et al. \cite{marinoni99}) also argues  that the  
early and late-type OLFs are very similar. NOG early-types show a flatter 
OLF than in CIG with brighter $M^*$. The NOG $M^*$ for elliptical galaxies 
is 0.3--0.4 brighter relative to both spirals and lenticulars while we find 
$M^*$ to be  very similar for our E and S0 subsamples (and less than for 
spirals). This again likely reflects the inclusion of an overluminous 
elliptical population found in the richer environments sampled by NOG 
but that are absent from AMIGA. NOG results for late-type galaxies show 
$M^*$ and $\alpha$ parameters consistent with the CIG. The steep 
$\alpha$ found by NOG ($-2.3$) is presumably  driven by local group dwarfs. 
In Paper I we also obtained a steeper $\alpha$ ($-1.3$ instead of $-0.8$) 
when galaxies within $V_{\rm R} < 1500$ km\,s$^{-1}$  were included although 
not as  steep as NOG. That may reflect an  underrepresentation of luminous 
spirals within $V_{\rm R} = 5500$ km\,s$^{-1}$  thus allowing $\alpha$ to 
drive the Schechter fit.  The $\alpha$--$M^*$ degeneracy makes 
this possible.  The type-specific OLFs given in NOG do not indicate 
any environmental discrimination.

The SDSS has been used to study the properties of a void sample (Rojas 
et al. \cite{rojas04}; Hoyle et al. \cite{hoyle05}). Void galaxies are found 
to be significantly bluer over a wide luminosity range. If one assumes that 
blue = late-type $\sim$ Sb--Sc then AMIGA is consistent with such a result.
SDSS found $M^*$ for their void sample to be one magnitude fainter than 
their ``wall'' sample. All or most of this difference can again be ascribed  
to the presence of a significant overluminous early-type population in 
the wall  sample. The $M^*$ and $\alpha$  parameters of the void sample 
are similar to  our complete sample values (Paper I). A closer comparison 
is not possible because, unless we misinterpret the  definition 
(i.e. fewer than three 
volume limited neighbors within a fixed radius), the selection does not 
exclude one-on-one interactions. An isolated pair sample like CPG 
(Catalog of Paired Galaxies; Xu \& Sulentic \cite{xu91}) would not be 
excluded from the void population because such pairs are found in regions of 
low density contrast. Evidence that: 1) components of such pairs are twice 
as bright as isolated galaxies of similar types and 2) interacting pairs 
comprise 10\% of the field 
galaxy population suggests that this contamination can confuse our 
interpretation of an environment signature in this kind of sample 
(see I/A=y sample in Table~\ref{olf}).

Stocke et al. (\cite{stocke04}) argue for a luminous ``fossil'' elliptical
population in the CIG in contradiction with an earlier study (Sulentic \& 
Rabaca \cite{sulentic94}). We disagree for two reasons: 1) problems with
morphologies and 2) misinterpretation of the OLF comparison sample. We
disagree with a number of their assigned E/S0 types and argue that CIG
57, 178, 284 (observed with Chandra as an E or S0), 417, 427, 430  
(see Fig.~\ref{imag}), 589, 640 and 690 are spirals while the most luminous 
object in their sample (CIG 701) is an interacting pair (I/A=y). They compare
$M^* = -20.0$ mag ($H_0$ = 70 km\,s$^{-1}$\,Mpc$^{-1}$) derived from their 
complete sample of $n=26$ ellipticals with a value obtained in the CfA1/CfA2 
survey  (Marzke et al. \cite{marzke94a}\cite{marzke94b}), 
 and find agreement which they interpret as evidence for
an overluminous elliptical population in the CIG. We suggest that one
must consider this result in the context of $M^*$ values for the spiral
population as well. Both the CfA spiral and S0 $M^*$ values are 0.5
magnitudes fainter than the E value. An elliptical population brighter
than spirals is typically what one finds when comparing populations
for a sample that includes galaxies in richer environments. 
All of this assumes that the CfA morphologies are reliable and, in this
context, we mention that CfA1 types come from multiple sources while
CfA2 types were taken from POSS\,I. A follow up SSRS2 survey (Marzke et
al. \cite{marzke98}) finds agreement between $M^*$ for the E/S0 and S 
subsamples in contradiction with Marzke et al. (\cite{marzke94b}),  
spirals being  brighter in SSRS2 than in CfA. 
The disagreement between the SSRS2 and CfA surveys has been ascribed to
errors in Zwicky magnitudes or to the earlier study sampling galaxies
in richer environments.  A detailed study of Zwicky magnitudes (Bothun
\& Cornell \cite{bothun90}) suggests that some spiral types (e.g. highly
inclined) were measured systematically fainter in the Zwicky
system. This would contribute towards making CfA spirals too faint
relative to spirals in SSRS2. The latter environmental explanation
does not seem likely since a higher environmental density
would tend to produce brighter ellipticals, rather than fainter spirals.

Our own complete sample of  $n =$ 27  ellipticals yields $M^* = -20.16$ mag
which is not very different from the Stocke et al. (\cite{stocke04}) 
value, however earlier we showed how we
can significantly dim $M^*$ for the elliptical sample with a reasonable
truncation of the luminosity range used for the OLF estimation.  
We also showed how the faintness of our E/S0 sample relative to our much
larger spiral sample is significant and robust. It makes no sense for AMIGA 
and CfA early-type OLFs to agree more closely than the spiral ones.
 Rather than comparing
$M^*$ values based on small samples of ellipticals we prefer the
following line of reasoning. The key point is that the
luminosity of an ellipticals population is much more sensitive to
environment than a corresponding spiral population.  $M^*$ is robustly
brighter in early vs. late-type comparisons.  AMIGA and, e.g., 2dFGRS
show that ellipticals become fainter in isolated environments until
their mean (or $M^*$) value equals, or even drops below, the spiral
value. Luminous spirals do not show this environmental sensitivity. A
sample that shows an early-type $M^*$ similar in brightness to a
late-type $M^*$ does not contain a luminous fossil elliptical
population.   We are speaking here about
samples of luminous galaxies (e.g. brighter than $-$18 or $-$19). Surveys
that go deeper can confuse this straightforward reasoning if a large
population of dwarf galaxies are included in an OLF calculation (as
was previously discussed for the 2dFGRS spiral sample). None of the
samples discussed here goes significantly deeper than AMIGA. The only
effect of dwarf galaxies on this comparison involves
inclusion/exclusion of local dwarfs.

\section{Discussion\label{sect6}}

CIG-based AMIGA sample is a magnitude limited sample with quite reasonable 
level of completeness from which a volume limited sample could be selected.
The main goal of AMIGA  is to maximize a sample of the most isolated  
galaxies in the local Universe, galaxies  in regions of both: 1) low galaxy 
surface density and 2) unaffected  by one-on-one interactions. 
The overall impression gained from the morphology survey is that a 
truly isolated massive galaxy may not exist. Typically 
luminous isolated galaxies often show an apparent excess of dwarf companions 
that are 4--5 magnitudes  fainter than the AMIGA galaxy (see also Reda et al.
\cite{reda04} and Smith et al. \cite{smith04}). In many of these cases (I/A=?
 should be the most obvious examples) the AMIGA galaxy shows clear signs of 
distortion (e.g. CIG 72 in Fig.~\ref{imag}) raising the question of what kind 
of damage a dwarf companion can do to a massive spiral. In the case of CIG 72 
we also find a Seyfert nucleus further emphasizing  the question (CIG 634 in 
Fig.~\ref{imag} is a LINER). Our recent detailed VLA \ion{H}{i} study  of 
CIG 96 (Espada et al. \cite{espada05}) faced the same question. In this case 
a bright spiral showing morphological and kinematic disturbance in an 
``isolated'' environment except for a dwarf companion that is: 
1) 4.8 magnitudes fainter and 
2) with less than 1\% of the \ion{H}{i} mass of the CIG primary.

AMIGA identifies two primary
populations of extremely isolated galaxies from the CIG:1) 14\% E--S0 and 2) 
63\% Sb--Sc. The overall CIG early/late ratio R = 0.14/0.86
compares to earlier values 0.17/0.83 (Nilson \cite{nilson73}) and 0.20/0.80 
(Gisler \cite{gisler80}).
A twenty percent overlap  with SDSS provides good confirmation 
of our POSS\,II based conclusions. That the CIG is one of the most spiral-rich 
samples is expected if the morphology-density relation 
(Postman \& Geller \cite{postman84}) extends to the most isolated 
regions of  the large scale structure. However, interpreted as an extremely 
isolated sample, the detection of a 
significant population of early-type galaxies in CIG raises special questions. 
The most extreme view is to argue that nurture is much more 
important than nature in determining galaxy morphology. The most extreme 
manifestation of this view would see a complete absence of 
early-type galaxies in extremely isolated environments interpreting all
ellipticals as merger products and  all lenticulars as a products of spiral 
harassment (Moore et al. \cite{moore96}). 
Claims have been made that the most isolated  samples contain no elliptical 
galaxies and few lenticulars  (e.g. Einasto \& Einasto \cite{einasto87})
 but they were based upon a  small sample of galaxies from local CIG 
components 1 and 2 summarized in Sect.~\ref{sect1}.  
POSS\,I-based classifications suggested an early-type 
CIG fraction as high as 0.25 while other studies (Saucedo-Morales \& Bieging 
\cite{saucedo01}; Stocke et al. \cite{stocke04}; and this paper) 
find 10--15\% almost evenly divided between E and S0 types. 
Thess typical AMIGA environments are where merging and harassment should 
have the lowest probability of occurrence. The modest luminosities of our 
E population supports the inference that these galaxies are not products of 
major mergers. OLF calculations have been discordant with claims for (Stocke 
et al. \cite{stocke04}) and against (Sulentic \& Rabaca \cite{sulentic94}) 
the existence of a luminous ``fossil'' elliptical population in the CIG. 
Stocke et al. (\cite{stocke04}) results were discussed in the previous 
section, and missclassified objects as well as an offset in $M^*$ of their 
comparison sample was argued to explain the disagreement with our results.
But the most important point, irrespective of the CfA comparison, is that 
late-type galaxies in the CIG are brighter than the elliptical galaxies 
in the CIG. Elliptical galaxies viewed as fossil ellipticals should be 
brighter than the population from which the mergers will be produced. The 
OLF study of Sulentic \& Rabaca (\cite{sulentic94}) ruled out a population of
fossil ellipticals from compact group mergers and our new results appear to 
almost rule out major mergers of CPG pairs or pairs  of CIG spirals. 

If not major mergers then what are the isolated ellipticals? If not stripped 
spirals then what are isolated lenticulars? A few CIG ellipticals have been 
studied in more detail (Marcum et al. \cite{marcum04}) with some showing 
normal red colors and a few unusually blue colors. CIG 164 and  870 are 
examples of this latter class and deserve higher resolution study. 
Both of these galaxies show unusually strong FIR emission for quiescent 
ellipticals. It is not clear that they are ellipticals at all but it is also 
possible that they represent recent mergers.  Early \ion{H}{i} studies of CIG 
lenticulars (Haynes \& Giovanelli \cite{haynes80}) suggested that some of them 
showed excess \ion{H}{i} content and spiral-like \ion{H}{i} profiles. 
Image analysis of CIG 83, variously classified as an E or S0 
(including POSS\,II) 
shows a weak but well defined spiral pattern after subtraction of a bright 
bulge component (Saucedo-Morales \& Bieging \cite{saucedo01}). Recent work 
also shows that the CIG S0s follow the radio-FIR correlation for spirals 
(Domingue et al. \cite{domingue05}). Since this is generally interpreted as 
a correlation driven by star formation one can ask if these  lenticulars, 
rather than being a product of harassment, are not some  kind of natural 
extension of the spiral sequence. Strong line emission, or early-type 
absorption, spectra are also not uncommon among CIG early-types (Stocke et al. 
\cite{stocke04}) further evidence that they may not be typical of their 
morphological classes. Assumming that a significant fraction of our E and S0 
populations are {\it bona fide} early types they may represent a primordial
population.

In environmentally mixed (e.g. clusters through voids) samples the shapes of 
the early and late-type OLFs are very similar. Past work discussed above, 
and references within, indicate that the luminosities of early-type galaxies 
are  more environmentally sensitive, with the $M^*$ parameter decreasing  
with decreasing environmental density. This can be interpreted as the major 
signature of nurture among the early-types. Spirals appear to be insensitive, 
or much less sensitive to environment with luminous spirals found everywhere 
(albeit less often in clusters). AMIGA is very helpful for interpreting 
results of other surveys that sample galaxies in a wider range of 
environments, since it is an extreme where effects of environmental 
nurture are minimized. An additional term in the environmental equation 
involves one-on-one interactions.  They  can produce multiwavelength 
enhancement signatures that cannot be distinguished from signatures driven 
by average environmental density. One-on-one interactions appear to be 
strongest in spirals while environmental density effects are strongest 
in early-types. Approximately 10\% of field galaxies are found in  close pairs 
(and perhaps 2--3\% in triplets and compact groups) that are often quite 
isolated. Strongly interacting pairs are even found  in voids (Grogin \& 
Geller \cite{grogin00}). This is potentially a large enough population that, 
unless adequately taken into account, can confuse or diminish OLF signatures 
connected to the average galaxy surface density. The AMIGA sample is the 
first large sample of isolated galaxies where both aspects of nurture 
are  being carefully monitored.

\section{Conclusions\label{sect7}}
The AMIGA sample is the largest local sample of extremely isolated and 
luminous ($-19 \ga M _{B{\rm -corr}} \ga -22$) 
galaxies. Our morphological revision
shows that it is dominated by: 1) a modest ($n =$ 139) early-type (E--S0)
population and 2) a dominant ($n =$ 637) late-type (Sb--Sc) population.
The sample is extreme because the spiral population is more luminous that 
the elliptical one, an effect seen only in isolated or void-like environments. 
AMIGA is trying to avoid two forms of ``nurture'': one-on-one interactions 
and galaxy environmental density. We have removed a sample of 32 obviously 
interacting pairs.  The next step involves evaluating the degrees of 
environmental density in our sample. One-on-one  interactions like the 
$n =$ 32 
AMIGA rejects  produce a maximum nurture  signature among late-type galaxies.
Environmental density produces the maximum signature in the early-type 
galaxies. The  former signature produces a multiwavelength enhancement while 
the latter a multiwavelength dimming which we see in  the low luminosity of the
AMIGA early-type population. 

The low luminosities of the AMIGA early-type population relative to: 1) the 
AMIGA spiral population and 2) early-type populations found in most surveys,
 is one of the most interesting results of this study. Environment appears 
to be the reason that we contradict claims that the early and late-type 
OLFs are very similar. The contradiction is due to the presence or absence 
of bright ellipticals in a sample and this depends on environmental density.
AMIGA appears to have found the most nurture-free population of luminous 
early-type galaxies.

\parindent 0pt

\begin{acknowledgements}
LV--M, GB, UL, DE, SL, SV and EG are partially supported by DGI (Spain) AYA 
2002-03338 and Junta de Andaluc\'{\i}a TIC-114 (Spain), with additional support
 by the Secretar\'\i a de Estado de Universidades e Investigaci\'on (GB). 
JWS is partially supported by MEC the spanish sabatical grant SAB2004-01-04.
Funding for the creation and distribution of the SDSS Archive has been 
provided by the Alfred P. Sloan Foundation, the Participating Institutions, 
the National Aeronautics and Space Administration, the National Science 
Foundation, the U.S. Department of  Energy, the Japanese Monbukagakusho, 
and the Max Planck Society. The SDSS Web site is
www.sdss.org. The SDSS is managed by the Astrophysical Research Consortium
(ARC) for the Participating Institutions. The Participating Institutions are 
the University of Chicago, Fermilab, the Institute for Advanced Study, 
the Japan Participation Group, the Johns Hopkins University, Los Alamos 
National Laboratory, the Max-Planck-Institute for Astronomy (MPIA), 
the Max-Planck-Institute for Astrophysics (MPA), New Mexico State University, 
Princeton University, the United States Naval Observatory, and the University
of Washington.
\end{acknowledgements}

\end{document}